\documentclass[twocolumn,aps,prl,showpacs,raggedbottom,nobalancelastpage,amssymb,superscriptaddress]{revtex4}

\usepackage{graphicx}
\usepackage{amsmath}
\usepackage{amssymb}
\usepackage{bm}
\usepackage{color}

\definecolor{lightgrey}{rgb}{0.7,0.7,0.7}

\newcommand{\rmd}{{\rm d}}
\newcommand{\rmi}{{\rm i}}

\newcommand{\up}{\uparrow}
\newcommand{\dn}{\downarrow}

\newcommand{\half}{{\textstyle{\frac{1}{2}}}}

\newcommand{\de}{\delta}
\newcommand{\De}{\Delta}

\newcommand{\ignore}[1]{\relax}

\newcommand{\lR}{{l_{\rm R}}}
\newcommand{\DeR}{{\De_{\rm R}}}

\begin{document}
\title{
Towards a dephasing diode: asymmetric and geometric dephasing}
\author{Robert S. Whitney}
\affiliation{Institut Laue-Langevin, 6 rue Jules Horowitz, B.P. 156,
         38042 Grenoble, France.}
\author{Alexander Shnirman}
\affiliation{Institut f\"{u}r Theoretische
Festk\"{o}rperphysik and DFG-Center for Functional Nanostructures
(CFN), Universit\"{a}t Karlsruhe, D-76128 Karlsruhe, Germany.}
\affiliation{
         Institut f\"ur Theoretische Physik, Universit\"at   Innsbruck, 
Technikerstr. 25,  6020 Innsbruck, Austria.}
\author{Yuval Gefen}
\affiliation{Department of  Condensed Matter  Physics, The Weizmann
         Institute of Science,  Rehovot  76100,  Israel.}

\date{March 25, 2008}
\begin{abstract}

We study the effect of a noisy environment on spin and charge
transport in ballistic quantum wires with spin-orbit coupling
(Rashba coupling). 
We find that the wire then acts as a dephasing diode, 
inducing very different dephasing of the spins of  
right and left movers.  
We also show how Berry phase (geometric phase) in a curved wire
can induce such asymmetric dephasing,
in addition to purely geometric dephasing.
We propose ways to measure these effects through spin detectors, 
spin-echo techniques,
and Aharanov-Bohm interferometry.
 \end{abstract}

\maketitle

A very promising idea for future 
(quantum or classical) information processing is ``spintronics''
\cite{Loss-book,review-by-Rashba,review-by-Loss-Marcus-Kouwenhoven,
review-spins-in-nanotubes},
where electrons' spins (not their charges) are 
used to encode information.
However spins do not obey the same conservation laws as charges;
charges do not change sign but spins can flip.
Current conservation 
enforces symmetries on {\it charge} transport (Onsager relations). 
For example two-terminal devices always have the same left-to-right and 
right-to-left conductance in the linear-response regime of
negligible interaction/charging effects (diodes do not exist 
without interactions).
By contrast, asymmetries between left-to-right and right-to-left
{\it spin}-transport can occur in the linear-response regime for
two-terminal devices, if there is spin-orbit coupling 
(Rashba or Dresselhaus).
{\it Coherence} is a crucial aspect of
quantum transport, so here
we investigate analogous asymmetries in the dephasing 
(decay of coherence) of spins.
We then study how Berry (geometric) phases --- present in curved wires 
\cite{Berry-in-wires} ---
modify such asymmetric dephasing \cite{wmsg2005}.

The coherence of a superposition of two spin-states
at the Fermi-surface is quantified in terms of the purity,
$P={\rm tr}[\hat \rho^2]$, where $\hat\rho$ is a $2\times2$ density matrix. 
A pure superposition has maximal purity, $P=1$, 
while an equal classical mixture has minimal purity, $P=1/2$.
We use the term ``dephasing diode'' for a two-terminal device
in which spin-superpositions of left movers experience very different 
dephasing from right movers.  
For an ideal dephasing diode, an electron
injected into the device from the left lead (a right mover)
in an equal coherent superposition of spin-states ($P=1$)
would emerge completely dephased ($P=1/2$),
while an electron
injected into the device from the right (a left mover)
in any superposition with $P=1$ would emerge at the left
without being dephased at all (still having $P=1$).
This dephasing could be observed by measuring 
either certain spin components of the current, 
or by a conventional current measurement in an Aharonov-Bohm (AB)
interferometer.  
In this letter, we provide illustrations of dephasing diodes with 
straight and curved ballistic wires.
For the latter, the Berry phase
gives a {\it geometry-induced} contribution to dephasing, 
whose sign depends on the curvature's sign.

Noise causes dephasing, and
real devices have many sources of noise
(thermal or quantum), including electron-electron and electron-phonon 
interactions.  However, a clear experimental 
observation of asymmetric dephasing requires
control of the noise-power
(seeing the asymmetry change with the noise-power).  
Thus we propose taking a wire with low intrinsic noise,
modelled by ballistic non-interacting electrons,
and applying {\it man-made} noise to the magnetic fields and gates. 
The response time of gates/magnets is typically longer than the
time-of-flight of electrons from source to detector along a 
ballistic wire \cite{slow-noise}. Hence we study the effect of 
extremely slow (man-made or intrinsic) noise 
on non-interacting electrons with spin-orbit
coupling.
For simplicity, here we consider only Rashba coupling,
${\bf B}_{\rm R} = \hbar (\hat p_x{\bf e}_y -\hat p_y{\bf e}_x)/ (m\lR)$,
where $\lR$ is the spin-precession length \cite{deriving-spin-orbit}, 
and magnetic fields are in units of energy.
Biasing a back-gate gives control over $\lR$ by modifying
the potential gradient  
along the $z$-axis \cite{Nitta97+Engels97}.
Noise can be applied to either the 
applied magnetic-field or the Rashba spin-orbit coupling (via 
a noisy voltage on the backgate).
By measuring any three orthogonal spin polarizations 
$\langle \sigma_i \rangle$, $i\in \{1,2,3\}$, 
one gets the purity 
$P=\half \big(1+\langle\hat \sigma_1\rangle^2
 +\langle\hat \sigma_2\rangle^2
 +\langle\hat \sigma_3\rangle^2\big)$~\cite{footnote:non-orth}.
Our qualitative predictions are summarized in
Table~\ref{table1}.  
Dresselhaus coupling yields similar results, but it is hard to control 
experimentally (i.e. not affected by the back-gate) 
so we do not consider it further here \cite{footnote:Rashba-Dressel}.

\begin{figure}
\includegraphics[width=8.5cm]{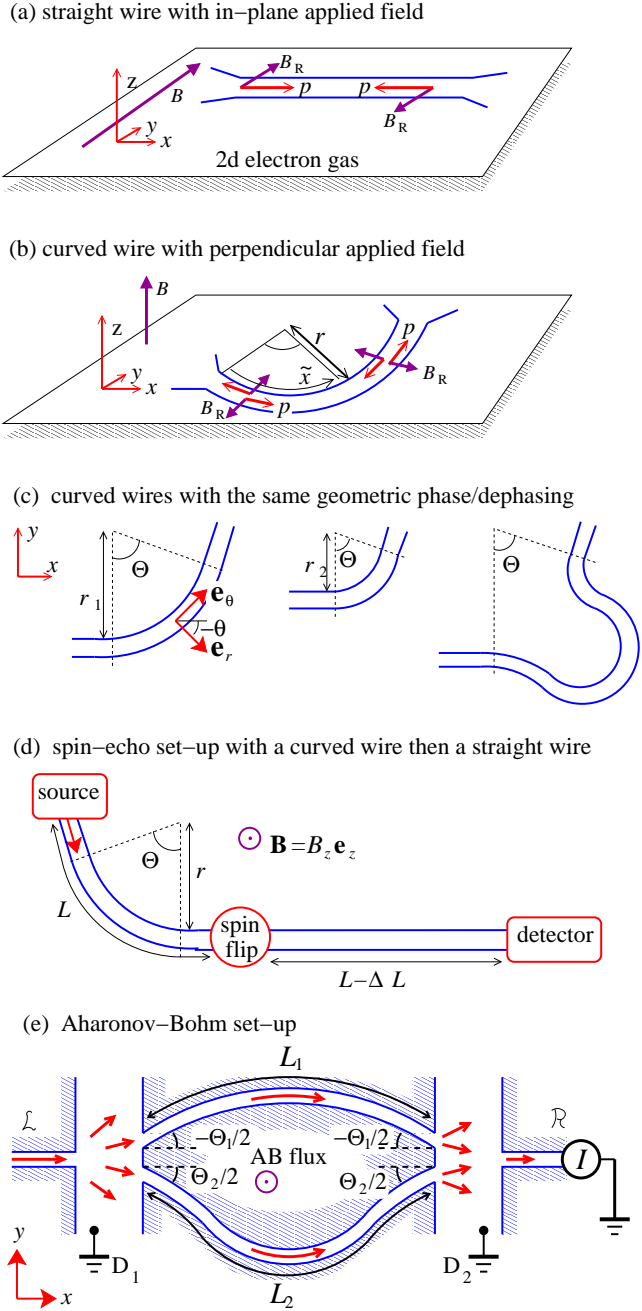}
\caption[]{\label{Fig:1}
The geometries we consider, with applied field, $B$ 
and Rashba field, $B_{\rm R}$. 
In (c), the BP changes sign for wires reflected 
in the $x$-axis ($\Theta \to -\Theta$). 
In (d) the source injects a superposition of $\up$ and $\dn$ eigenmodes,
and the spin-flip takes $\up \leftrightarrow \dn$.
In (e) 
spin-polarized electrons are injected at ${\cal L}$. 
Some are detected at ${\cal R}$, but most escape 
into $D_{1,2}$.
}
\end{figure}

\begin{figure}
\includegraphics[width=7.5cm]{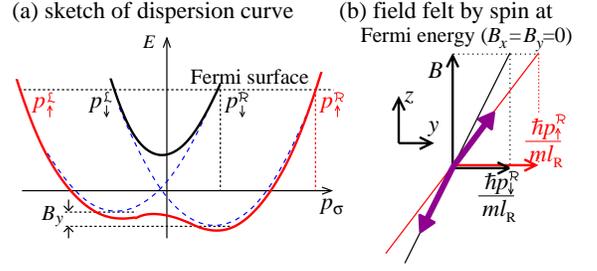}
\caption[]{\label{Fig:2}
(a) A sketch of the dispersion curve for $\hat{\cal H}$ in  
Eq.~(\ref{eq:H}); dashed parabolas have $B_x=B_z=0$ \cite{Governale02b}.  
(b) The effective field
felt by spin-states at $E_{\rm F}$ (for $B_x=B_y=0$).
Since $p_\up^{\cal R} > p_\dn^{\cal R}$, 
the two states have non-orthogonal spins
(their overall orthogonality is due to 
$p_\up^{\cal R}\neq p_\dn^{\cal R}$).
}
\end{figure}

{\bf Asymmetric dephasing in a straight wire}
(see Fig.~\ref{Fig:1}a).
We neglect the contribution of the motion across the wire 
($y$-direction) \cite{footnote:H_y}, so the 
Hamiltonian for an electron in the lowest mode of 
the ballistic wire is
\begin{eqnarray}
\hat{\cal H} = 
(2m)^{-1} \hat p_x^2 + E_0  
-\hbar(m\lR)^{-1} \hat p_x \hat \sigma_y - \half {\bf B}\cdot\hat{\bf \sigma}, 
\label{eq:H}
\end{eqnarray}
where $E_0$ is the transverse mode's energy.
Hence the eigenmodes have momenta, $p_\sigma$, given by
\begin{eqnarray}
0&=& p_\sigma^4 -[2p_0^2 +4(\hbar/\lR)^2] p_\sigma^2  
   + [2\hbar mB_y/\lR]   p_\sigma 
\nonumber \\
& & \qquad 
+ p_0^4 - (mB)^2,   
\label{eq:quartic-p}
\end{eqnarray}
where $B=|{\bf B}|$ and $p_0 = [2m(E_{\rm F}-E_0)]^{1/2}$.
Here $\sigma=\pm 1=(\up,\dn)$ is the spin-state, 
orientated along the effective field
$(B_x,2\hbar(m\lR)^{-1}p_\sigma +B_y,B_z)$,
see Fig.~\ref{Fig:2}.

\begin{table*}
\begin{tabular}{|r||c|c|c|c|} 
\hline
Wire \ \ \ 
& $\lR$-noise (cf. Eq.~(\ref{eq:H}))   & \, $B_z$-noise  
& \ \ $B_y$-noise \ \  & \ \ $B_x$-noise\ \  
\\ 
\hline\hline 
\ \ Straight, finite $B_y$ \ \ 
&  sym. $+$ weak asym. & none &  sym.  &   none 
\\ 
\hline
\ Straight, finite $B_z$ \ \ 
& sym. & sym. & sym. &    sym.  
\\
\hline
\ Straight, finite $B_x$ \ \ 
& sym. & sym. & sym. & sym.  
\\
\hline
\ Curved, finite $B_z$ \ \ 
& sym. $+$ weak geom.-induced asym.\  
&  sym. $+$  weak geom.-induced asym.
\\
& $+$ weak purely geometric sym.&  $+$ weak purely geometric sym. 
\\
\cline{1-3}
\ Curved $+$ spin-echo  \ \ 
 & \ sym. $+$ strong geom.-induced asym. \ 
& \ sym. $+$ strong geom.-induced asym.  \
\\
& $+$ strong purely geometric sym. 
& $+$ strong purely geometric sym.\ 
\\
\cline{1-3}
\ AB set-up, \ \ 
& \ sym. $+$ strong geom.-induced asym.\ & sym. $+$ weak asym. 
\\
finite $B_z$ \ \  
& $+$ strong purely geometric sym. & \ $+$ weak geom.-induced asym.\  
\\
& & $+$ weak purely geometric sym. 
\\
\cline{1-3}
\end{tabular}
\caption{\label{table1}
Contributions to dephasing for the various systems:
``none'' 
indicates the absence of dephasing 
~\cite{footnote:other-noise};
``sym./asym.'' 
indicate that the terms
are symmetric/asymmetric under right movers $\leftrightarrow$ left movers.
Terms labelled ``strong''(``weak'') are of similar size to 
(much less than) the main symmetric term, 
for typical experimental parameters. 
``Geom.-induced'' and ``purely geometric'' indicate terms which vanish for 
a straight wire, the former go like $L\Theta$ while
the latter go like $\Theta^2$.
}
\end{table*}

Let us first consider $B_x=B_z=0$, 
then the solutions of Eq.~(\ref{eq:quartic-p}) are 
$p_\sigma^{\cal R,L} = (\hbar/\lR) \sigma \pm [p_0^2+(\hbar/\lR)^2 + mB_y \sigma]^{1/2}$
with the upper and lower sign for ${\cal R}$ and ${\cal L}$ movers, 
respectively.
We assume $p_0^2 > mB_y$, 
then an ${\cal R}$ or ${\cal L}$ mover in a superposition of two spin-states,
$\up$ and $\dn$ (e.g. spin in the $x$-direction), 
acquires a phase difference of  
$\Phi^{\cal R,L} = (|p_\up^{\cal R,L}|-|p_\dn^{\cal R,L}|)L/\hbar$
between the two spin-states when traversing the wire.
There is no energy-term because both
states have the same energy. 
Given $p_\sigma^{\cal R,L}$ we have
$\Phi^{\cal R,L} 
= \big(\Lambda_{B_y}^{-1} - \Lambda_{-B_y}^{-1} \pm 2\lR^{-1} \big)L$
where the length scale 
$\Lambda_B\equiv \hbar [p_0^2 +(\hbar/ \lR)^2 + mB]^{-1/2}$. 
Noise smears this phase difference causing dephasing.
We consider Gaussian-distributed noise in ${\bf B}$ and $\lR$ which is 
much slower than the time-of-flight between source 
and detector.
We use 
$\langle \cdots \rangle$ to indicate taking the expectation value, 
{\it and} averaging over $\de B_i, \de \lR$
with weight
$ \exp \! \big[
-\half\sum_i(\de B_i/\De_i)^2 -\half(\de \lR/\DeR)^2 
\big] $. 
Assuming that the noise is weak, $ \De_i \ll B_i$ and $\DeR \ll \lR$, 
we can expand $\Phi^{\cal L,R}$ 
to first order in $\de B_i$ and $\de \lR$. 
Defining $\hat\sigma_1$ as along the eigenbasis 
(here the $y$-axis), 
we have 
\begin{eqnarray}
& & \hskip -8mm \langle \hat \sigma_{2,3}\rangle 
\ \propto \ \Big|\langle\exp[\rmi \Phi^{\cal R,L}] \rangle \Big|
\nonumber \\
&=&  
\exp \big[-{\textstyle {1\over 8}}
\big(L(\Lambda_{B_y}+\Lambda_{-B_y})m\De_B/\hbar^2\big)^2 \big] 
\nonumber \\
& & \times 
\exp \big[- \half \big(
L(\Lambda_{B_y}-\Lambda_{-B_y} \pm 2\lR)\De_{\rm R}/\lR^3\big)^2 \big]
\label{eq:B_y-Phi-averaged} 
\end{eqnarray} 
This yields a Gaussian decay of the purity with $L$,
i.e. with the time-of-flight along the wire \cite{footnote:other-noise}.
Such Gaussian decays are typical 
of slow noise (inhomogeneous broadening) \cite{inhomogeneous-broadening}.
Noise 
in $B$ dephases the spin of ${\cal R}$ and ${\cal L}$ movers
   in the same way (symmetric).  
However noise in $l_{\rm R}$ dephases 
${\cal R}$ movers 
differently from ${\cal L}$ movers [upper vs. lower sign in Eq. (3)].
This asymmetry can only be large if  $|\Lambda_{B_y} -\Lambda_{-B_y}|\sim 2l_{\rm R}$.

In contrast, for $B_y=0$,
Eq.~(\ref{eq:quartic-p}) is
a quadratic equation for $p_\sigma^2$.
For 
every ${\cal L}$ mover with momentum $p_\sigma$ 
there is a ${\cal R}$ mover with momentum $-p_\sigma$.
In this case $\Phi^{\cal R}=\Phi^{\cal L}$, so there is 
no asymmetry in dephasing.


{\bf Geometric dephasing in a curved wire,}
see Fig.~\ref{Fig:1}b.
The electron travelling along the wire is subject to a spatially varying
effective field (${\bf B}+{\bf B}_{\rm R}$, where ${\bf B}=B_z{\bf e}_z$).
We go to cylindrical coordinates $(r,\theta,z)$ \cite{MMK02}
dropping terms that go like $\lambda_{\rm F}/r$. 
Transforming the spinor using
$\hat{\cal U}_\theta = \exp [\rmi \half\theta \hat \sigma_z] $
gives the Hamiltonian
\begin{eqnarray}
\hat{\cal H}_{\rm cyl} 
&=& 
{\hat p_\theta^2\over 2m} + E_0 
-{\hbar \hat p_\theta \over m \lR} \hat \sigma_r 
- \left(B_z + {\hbar \hat p_\theta \over 2mr }\right) \hat \sigma_z \quad
\label{eq:H-cyl}
\end{eqnarray}
where the $\theta$-dependence of the frame causes the 
$\hat p_\theta \hat \sigma_z$ term.
The eigenmodes' momenta, $p_\sigma\equiv p_\theta(\sigma)$, are given by
Eq.~(\ref{eq:quartic-p}) with $B_y=B_x=0$ and $B_z$ replaced by  
$(B_z + (2mr)^{-1} p_\sigma)$.
We next assume a large radius of curvature,
$r \gg \hbar p_0[(\hbar p_0/\lR)^2+(mB_z/2)^2]^{-1/2}$ 
(so the frame's angular velocity
$\ll$ precession rate in the effective field)
and expand  
$p_\sigma = \pm |p_\sigma^{\infty}| +\hbar c_\sigma r^{-1}$,
where $p_\sigma^{\infty}$ is the momentum for $r=\infty$.
To order $1/r$, we obtain 
$c_\sigma = \sigma \cos \kappa$
where we define $\kappa$ as the angle between the $z$-axis 
and an effective field $(0, 2\hbar[p_0^2 +(\hbar/\lR)^2]^{1/2}/(m\lR),B_z)$.
This is readily generalized to other wire shapes (cf.~Fig.~1c),
with $r$ varying along the wire coordinate, $\tilde x$,
($r(\tilde x)$ to be kept large). Noting that 
$(c_\up - c_\dn)=2c_\up$,
we obtain
\begin{eqnarray}
\Phi^{\cal R,L} 
 &=& \hbar^{-1}
\int \big ( |p_\up^{\cal R,L}|-|p_\up^{\cal R,L}| \big) r(\tilde x) \rmd 
\theta(\tilde x) 
\nonumber \\
&=&
\big(|p_\up^{\infty}| - |p_\dn^{\infty}|\big) L/\hbar
\, \pm \, 2c_\up \Theta \qquad 
\label{eq:Phi-curved}
\end{eqnarray}
The first term in $\Phi^{\cal R,L}$ goes like the wire length, $L$,
(i.e.~proportional to the time-of-flight along the wire)
and is hence a {\it dynamic} phase.  
The second term is proportional to the total change in angle, $\Theta$, 
is independent of the 
time-of-flight and is thus a {\it Berry} (geometric) phase;
it is the same for all wires in Fig.~\ref{Fig:1}c.
As $\Theta$ is a directed angle, $\Theta\to -\Theta$ means a curve in the 
opposite sense (i.e.~clockwise$\to$counter-clockwise).
Introducing noise (in $B_z$ and $\lR$) now adds the factor 
$\big[\de\lR\,{\rmd\over \rmd \lR} + \de B_z\,{\rmd\over \rmd B_z}\big] 
[(|p_\up^\infty| - |p_\dn^\infty|)L/\hbar \pm 2c_\up \Theta]$
to $\Phi_{\cal R,L}$.
Averaging over this noise as before yields decay (dephasing)
of the purity 
with the exponent
\begin{eqnarray}
   -\half \Big(
    \big[ \De_{\rm R} {\textstyle{\rmd\over \rmd \lR}} 
         +\De_{B_z}{\textstyle{ \rmd\over \rmd B_z}} \big] 
   \big[\big(|p_\up^{\infty}| - |p_\dn^{\infty}|\big) L/\hbar 
   \pm 2c_\up\Theta \big] 
   \Big)^2 \ \ 
\label{eq:dephase-curved}
\end{eqnarray}
The $\sim L^2$ and $\sim \Theta^2$ terms in the exponential
(dynamic and purely geometric terms respectively) are both unchanged
under ${\cal R} \leftrightarrow {\cal L}$.
The cross term, $\sim L\cdot \Theta$, 
(a mixed dynamic-geometric term)
 changes sign under ${\cal R} \leftrightarrow {\cal L}$,
causing a (geometry-induced) left-right asymmetry.

Coherent oscillations carry the phase of
Eq.~(\ref{eq:Phi-curved}), their amplitude decays with the exponent in
Eq.~(\ref{eq:dephase-curved}). 
Ambiguity in choosing the measurement axis 
(any axis perpendicular to the axis of the eigenbasis)
causes ambiguity in the 
phase (know as gauge-dependence), but not the amplitude.  
Thus the geometric contribution to dephasing is 
gauge-{\it independent} even when the BP is not \cite{wmsg2005}.

{\bf Spin-echo.} 
To maximize the geometric or asymmetric effects one may use a 
spin-echo technique, sketched in Fig.~\ref{Fig:1}d.
If $\Theta=\Delta L=0$, the wires left and right of the spin-flipper
are identical, and any spin-component acquires opposite phases 
before and after the spin-flip.
Thus for non-zero $\Theta$ and $\Delta L$, the phase is
given by Eq.~(\ref{eq:Phi-curved}) with $L$ replaced by $\De L$.
Dephasing is given by Eq.~(\ref{eq:dephase-curved}) with $L \to \De L$.
Varying $\De L$ changes the relative size of the
$\De L^2$, $\De L\,\Theta$ 
and $\Theta^2$ contributions to dephasing.
For $\De L=0$, there is only purely geometric dephasing.
Asymmetry is maximized for a small $\De L$, 
such that the $\De L^2$ and $\De L\,\Theta$ 
terms are similar in size.

{\bf Aharonov-Bohm interferometer.}
Measuring the flux-sensitive current through such an interferometer
(Fig.~\ref{Fig:1}e)
allows us to study dephasing \cite{AB-dephasing}, 
while avoiding the need to measure 
spin-components of the current.
We generate right or left movers via a voltage bias to either the 
${\cal L}$ or the ${\cal R}$ lead.
Asymmetry in dephasing will manifest itself as a difference in
the visibility (magnitude of the AB oscillations in the current)
between right and left movers,
We assume that the multi-terminal (open) interferometer 
is sufficiently open that no higher windings around it occur.
This also 
avoids the symmetry constraints imposed by the two-terminal
Onsager-B\"uttiker relations.
We consider injected electrons which are spin-polarized
along ${\bf B}=B_z{\bf e}_z$,
thus the phase difference between the two paths is
$
\Phi^{\cal R,L} = \pm\Phi_{\rm AB} + |p_\up^\infty| L/\hbar
\pm c_\up \Theta
$
where $L=L_2-L_1$ and $\Theta=\Theta_2-\Theta_1$.
The Aharonov-Bohm phase $\Phi_{\rm AB} = -eB_z{\cal A}/(g \mu_{\rm B}\hbar)$ 
where ${\cal A}$ is the area enclosed by the paths.
The current at the detector is 
\begin{eqnarray}
|I^{\cal R,L}| = |I_1| + |I_2| + 2|I_1 I_2|^{1/2} 
A^{\cal R,L} \cos \Phi^{\cal R,L}
\label{eq:I_AB}
\end{eqnarray}
where $I_i$ is the part of the current in arm $i$
which enters the detector lead.
In the absence of noise $A^{\cal R,L}=1$.
For $I_1=I_2$, 
the visibility of the AB oscillations is maximal.
Averaging over noise in $B_z$ and $\lR$,
we find that 
\begin{eqnarray}
  A^{\cal R,L} \!\! &=& \! \exp\Bigg[-\half
    \De_{B_z}^2 \left(
    {e {\cal A} \over g\mu_{\rm B}\hbar} 
    \pm {L\over \hbar} {\rmd |p_\up^\infty| \over\rmd B_z} 
    +  \Theta {\rmd c_\up\over\rmd B_z} 
\right)^2
\nonumber \\
& & \qquad 
-\half
    \De_{\rm R}^2 \Big(  
    \pm {L\over \hbar} {\rmd |p_\up^\infty| \over\rmd \lR} 
    +  \Theta {\rmd c_\up \over\rmd \lR} 
\Big)^2 \Bigg].
\label{eq:dephase-AB}
\end{eqnarray}
This is asymmetric, 
with the upper (lower) sign for 
${\cal R}$ (${\cal L}$) movers.
There are contributions to dephasing due to the curvature,
$\sim L\Theta$ (asymmetric) and 
$\sim \Theta^2$ (symmetric).

{\bf Discussion.}
Quantum wires in Ga$_x$In$_{1-x}$As/InP have 
$\lR \simeq 5\lambda_{\rm F} \simeq 200$nm \cite{Schapers06}.
Then the ratio of asymmetric to total dephasing in a straight-wire
(cf.~Eq.~(\ref{eq:B_y-Phi-averaged}))
will be small, equal to $2\lambda_{\rm F}^2/(\lR l_B) \sim 8\%$, 
where $l_B = v_{\rm F}\hbar/B \sim \lR$. 
However for spin-echo set-ups with $\De L\sim l_{\rm R}$, 
this ratio can be of order one 
(cf.~Eq.~(\ref{eq:dephase-curved}) and thereafter).
Then the dephasing is not very strong, 
nonetheless we estimate that one can tune $\De L$
such that
${\cal R}$ movers lose at least $50\%$ of their coherence
(purity $P < 3/4$), while ${\cal L}$ movers are not dephased.
For the AB set-up with $l_{\rm R}$-noise (cf.~Eq.~(\ref{eq:dephase-AB})), 
the situation
is the same as for the spin-echo (with $L$ playing the role of $\Delta L$),
while for $B_z$-noise the ratio of asymmetric to total dephasing
(in the exponent) is tiny $\sim \lambda_{\rm F}L/{\cal A}$.

To summarize,
we have analyzed the effect of noise on current-carrying electrons,
both spinful and spin-polarized,
subject to a spin-orbit interaction. 
The dephasing may be measured through the  
spin-components of the emerging electrons,
or the visibility of AB oscillations.
 We have demonstrated both geometric dephasing and left-right asymmetry
in dephasing.  The noise studied here fluctuates at a pace
slower than the time-of-flight, leading to dephasing with the 
exponential of $\sim L^2$, $\sim L\cdot\Theta$, $\sim \Theta^2$.
By contrast, for faster fluctuations 
(studied for a spin or qubit in a 
time-dependent field~\cite{wmsg2005}), 
the dephasing would be exponential with $L$, $\Theta$,
see also \cite{SanJose06,wg2003+wmsg-conf-proc,Wallraff-expt07}.
Other
${\cal L} \leftrightarrow {\cal R}$ asymmetries occur in real devices,
however those discussed here have a unique signature;
they are controlled by the noise-power.

We thank Yu.~Makhlin and G.~Sch\"on for very useful discussions.
This work was supported by
Swiss NSF,
FG Schwerpunktprogramm 1285 "Halbleiter Spintronik",
EU Transnational Access program (RITA-CT-2003-5060965), 
US-Israel BSF, Minerva Foundation and 
Albert Einstein Minerva Center for Theoretical Physics. 



\bibliographystyle{apsrev}

\begin{thebibliography}{9}
\bibitem{slow-noise}
The gate RC times $\gtrsim 10^{-6}{\rm s}$ 
(although they may be engineered to be shorter), while
the time-of-flight along ballistic wires
($L<0.1{\rm mm},v_{\rm F}\sim 10^5{\rm m/s}$) is $\lesssim 10^{-9}{\rm s}$.
\bibitem{footnote:non-orth}
This equation for $P$ is approximate
when Fermi-surface states
have slightly non-orthogonal spins, Fig~\ref{Fig:2}b.
\end{thebibliography}

\end{document}